\documentclass[letters,10pt]{IEEEtran}

\usepackage{bm} 
\usepackage[latin9]{inputenc}
\usepackage{amsmath}
\usepackage{amssymb}
\usepackage{pdflscape}
\usepackage[table]{xcolor}
\usepackage{longtable}
\usepackage{tabularx}
\usepackage{multirow}
\usepackage[dvips]{graphicx}
\usepackage{epsfig}
\usepackage{algorithmic}
\usepackage{algorithm,caption}
\usepackage{romannum}
\usepackage{booktabs}
\usepackage{array}
\usepackage{float}
\usepackage{cite}
\usepackage{mdwlist}
\usepackage{url}
\usepackage{epstopdf}
\usepackage{tikz}
\usetikzlibrary{tikzmark,fit,shapes.geometric,positioning,calc}
\ifCLASSOPTIONcompsoc
  \usepackage[caption=false,font=normalsize,labelfont=sf,textfont=sf]{subfig}
\else
  \usepackage[caption=false,font=footnotesize]{subfig}
\fi
\usepackage{balance}
\usepackage{multicol}

\DeclareMathOperator*{\argmin}{arg\,min}

\newcommand{\beff}{{\mathbf{f}}}

\newcommand{\be}{{\mathbf{e}}}

\newcommand{\bzero}{{\mathbf{0}}}

\newcommand{\bhf}{\mathbf{{\hat{f}}}}

\newcommand{\bx}{{\mathbf{x}}}
\newcommand{\btx}{{\mathbf{\tilde{x}}}}
\newcommand{\by}{{\mathbf{y}}}
\newcommand{\bR}{{\mathbf{R}}}
\newcommand{\bhR}{{\mathbf{\hat{R}}}}
\newcommand{\bS}{{\mathbf{S}}}

\newcommand{\bI}{{\mathbf{I}}}

\newcommand{\bphi}{{\mbox{\boldmath$\phi$}}}

\newcommand\Tstrut{\rule{0pt}{2.6ex}}         

\tikzstyle{decision} = [diamond, aspect=2, draw, thick, fill=yellow!15!white, 
	    text width=0.2\linewidth, align=center, inner sep=0pt] 
\tikzstyle{block} = [rectangle, draw, rounded corners=3pt, fill=cyan!15!white, text width=0.25\linewidth, thick, align=center, rounded corners, minimum height=0.5cm]
	
\tikzstyle{box} = [rectangle, draw,fill=white, text width=0.05\linewidth, thick, align=center]

\begin{document}


\title{A Method with Lower-than-ML Threshold for Frequency Estimation of Multiple Sinusoids}

\author{P. Vishnu and C.S. Ramalingam
\thanks{P. Vishnu and C.S. Ramalingam are with the Department of Electrical Engineering, IIT Madras. Email: ee12d038@ee.iitm.ac.in, csr@ee.iitm.ac.in.} 
}


\maketitle

\begin{abstract}
Estimating the frequencies of multiple sinusoids in the presence of AWGN and when the data record is short is commonly accomplished by subspace-based methods such as ESPRIT, MUSIC, Min-Norm, etc.  These methods do not assume that the data are zero outside the observation interval.  If we assume otherwise, the threshold SNR is lowered significantly, but the price paid is unacceptable bias.  Among all known unbiased estimators, the maximum-likelihood estimator (MLE) has the lowest threshold, but is computationally the most expensive.  We propose a new algorithm that carries out, when needed, (i) zero-padding, and (ii) removal and re-estimation.  These added steps result in a threshold SNR that is \textit{lower than that of the MLE} for the examples considered herein, viz., noisy signals containing sinusoids with random parameters and up to five components.  The maximum improvement in threshold was $10\,$dB for the two-sinusoid case.  The bias of the estimates is also either equal to or lower than MLE's.  Unlike the MLE, the proposed method is very much computationally feasible.
\end{abstract}


\section{Introduction}
In this paper we revisit the classic and well-studied problem of estimating the parameters of multiple sinusoids in the presence of AWGN \cite{kaySpectralBook,stoica-moses-2005,marple-book-2019}.  The noisy observed data, consisting of $p$ complex sinusoids, can be modeled as
\begin{equation}
x[n] = \sum_{l=1}^p v_l\,e^{j2\pi f_l n} + w[n]\quad n=0, 1, \ldots, N-1
\label{eq:data model}
\end{equation}
The complex amplitudes $v_l$ $(=|v_l|e^{j\phi_l})$'s and the
frequencies $f_l$'s are the unknowns; $p$ is assumed to be known.  Furthermore, we assume that the data record is short, containing frequency components that are spaced below the resolution limit of Fourier-based methods, i.e., less that $1/N$.

If the $\phi_l$ are assumed to be random and uniformly distributed in the interval $[0, 2\pi)$, the autocorrelation sequence (ACS) has the form,
\begin{equation}
r_{xx}[k] = \sum_{l=1}^p P_l\,e^{j2\pi f_l k} + \sigma^2\,\delta[k]
\end{equation}
where $P_l = |v_l|^2$ and $\sigma^2$ is the noise variance.  The eigenstructure of the associated $M\times M$ autocorrelation matrix $\bR_{xx}$ is exploited by the so-called subspace-based methods \cite{Schmidt1986MUSIC,barabell1983RootMusic,Roy1986ESPRIT,kumaresan-82b,kumaresan1983doaEstimation}.  Since the true ACS is not known, the $r_{xx}[k]$ are \textit{estimated} from the given \textit{finite} data set.  If we assume that observed sequence $x[n]$ is zero outside the observation window, the estimated ACS is labeled as the ``autocorrelation estimate''; if, however, no assumption about $x[n]$ is made outside the observation window, the estimated ACS is termed as the ``covariance estimate'' (these terms are commonly used in speech processing literature \cite{rabiner-2011}).

The maximum-likelihood estimator (MLE) \cite{kaySpectralBook} has the lowest threshold SNR (the SNR below which the variance of the estimates registers a sharp increase) among all known methods.  The disadvantage is that the MLE is computationally the most burdensome method.  Subspace-based methods are a good compromise between computational burden and reasonably low (but higher-than-ML) threshold SNR.  Hitherto, all the subspace-based methods have used the ``covariance'' assumption because the ``autocorrelation'' assumption yields estimates having an unacceptable bias even in the noiseless case.

Some recent contributions to the sinusoidal frequency estimation are \cite{bhaskar-tang-2013,mamandipoor-ramasamy-2016,Selva-2017,Ye-Aboutanios-2017,Djukanovic-Bugarin-2019, vishnu-csr-2018b}.  In \cite{mamandipoor-ramasamy-2016} the problem was viewed as one of finding a sparse approximation of the signal with an infinite dimensional dictionary of sinusoids.  For this method, one needs a minimum frequency separation of $2.52/N$ for accurate estimation, which is higher than the frequency separation considered in this work.  In \cite{Ye-Aboutanios-2017} the assumption is that the components are resolved and the goal is to estimate their frequencies cheaply yet accurately.  In \cite{Djukanovic-Bugarin-2019}, the authors assume that sinusoids are at least one DFT bin apart.  Since the methods in \cite{Ye-Aboutanios-2017,Djukanovic-Bugarin-2019} are Fourier based, they are not suitable for estimating frequencies with separation less than $1/N$.  In \cite{vishnu-csr-2018b}, we proposed improved frequency estimation method for the closely spaced sinusoids cases (i.e., frequency separation less than $1/N$).  The method in \cite{Selva-2017} is an estimation-detection method, i.e., the number of sinusoids present are detected and their frequencies estimated.  A ``residual periodogram'' is used for detecting new frequencies until all of them have been found.  In our case we focus on short data records and closely spaced components, assuming that their number is known.

In this work we propose a method that uses the ``autocorrelation'' assumption as part of it.  Further processing is carried out to reduce its bias.  For the examples considered herein, our algorithm gives unbiased estimates and leads to thresholds that are lower than MLE's, while still being very much computationally tractable.

\section{Motivation}
\label{sec:motivation}
Consider the well-known two-sinusoid example \cite{KTmethod1982Undamped} with $N = 25$, $f_1 = 0.52$, $f_2 = 0.5$, $|v_1| = |v_2| = 1$, and $\phi_1 - \phi_2 = 0$.  Conventional ESPRIT uses the ``forward-backward approach'' to estimate $\bhR_{xx}$ (size $M \times M$) by making no assumption about $x[n]$ outside the observation window.  On the other hand, if we assume that the given data are zero outside the observation window, we can use the zero-padded data for frequency estimation; we designate this method as ESPRIT-AC.  It is easy to see that ESPRIT-AC is nothing but the conventional ESPRIT operating on $\by = [\bzero \;\bx^T \;\bzero]^T$ (where $\bzero$ is of size $1 \times M$ and $\bx = (x[0]\;\;x[1]\;\;\ldots\;\;x[N-1])^T$).

\begin{figure}[t!]
\centering
\subfloat[ESPRIT]{\includegraphics[width = 1.6in,trim={0 0 0 
0.01cm},clip]{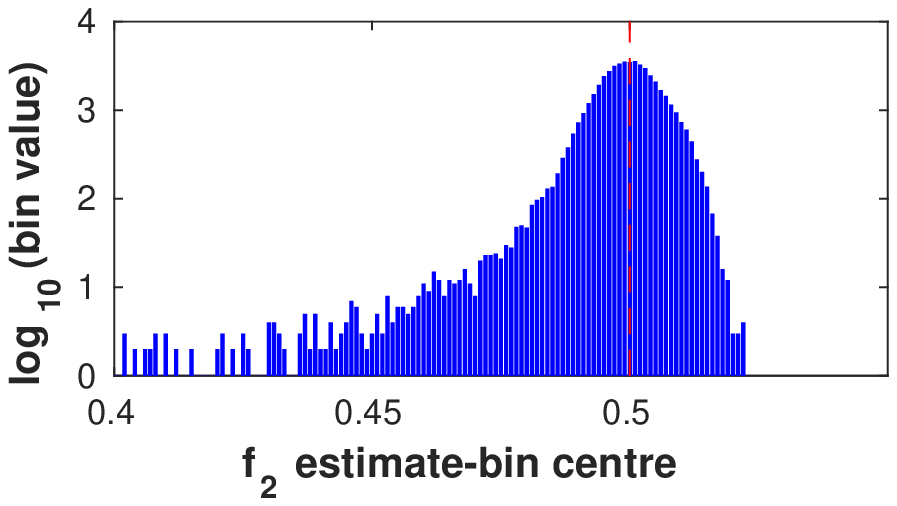}}$\,$
\subfloat[ESPRIT-AC]{\includegraphics[width = 1.6in,trim={0 0 0 
0.01cm},clip]{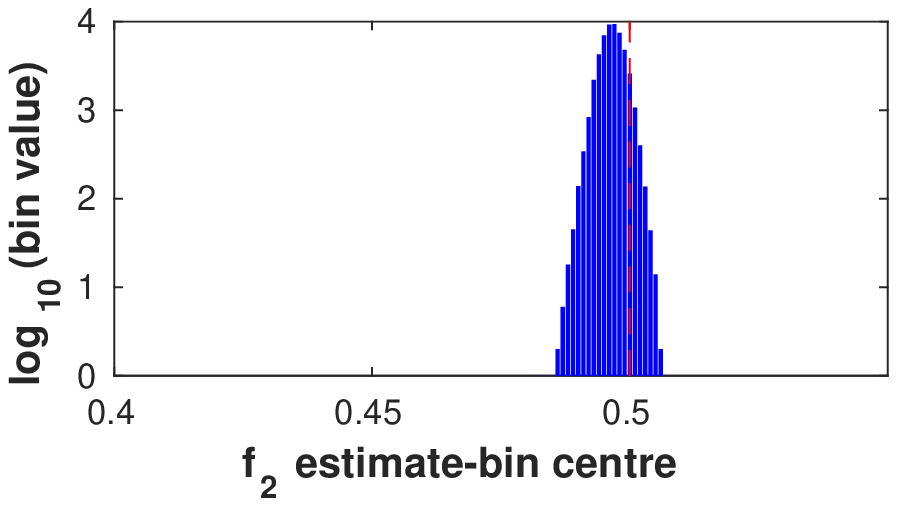}}
\caption{Histogram of $f_2$ obtained using ESPRIT and ESPRIT-AC for the two-sinusoid example ($f_1 = 0.52$, $f_2 = 0.5$, $\phi_1 - \phi_2 = 0$) at 5dB SNR (based on 50k trials).}
\label{fig:hist-esprit}
\end{figure}
\begin{figure}[h!]
\centering
\subfloat[MSE]{\includegraphics[width = 1.45in,trim={0 0 0 
0.01cm},clip]{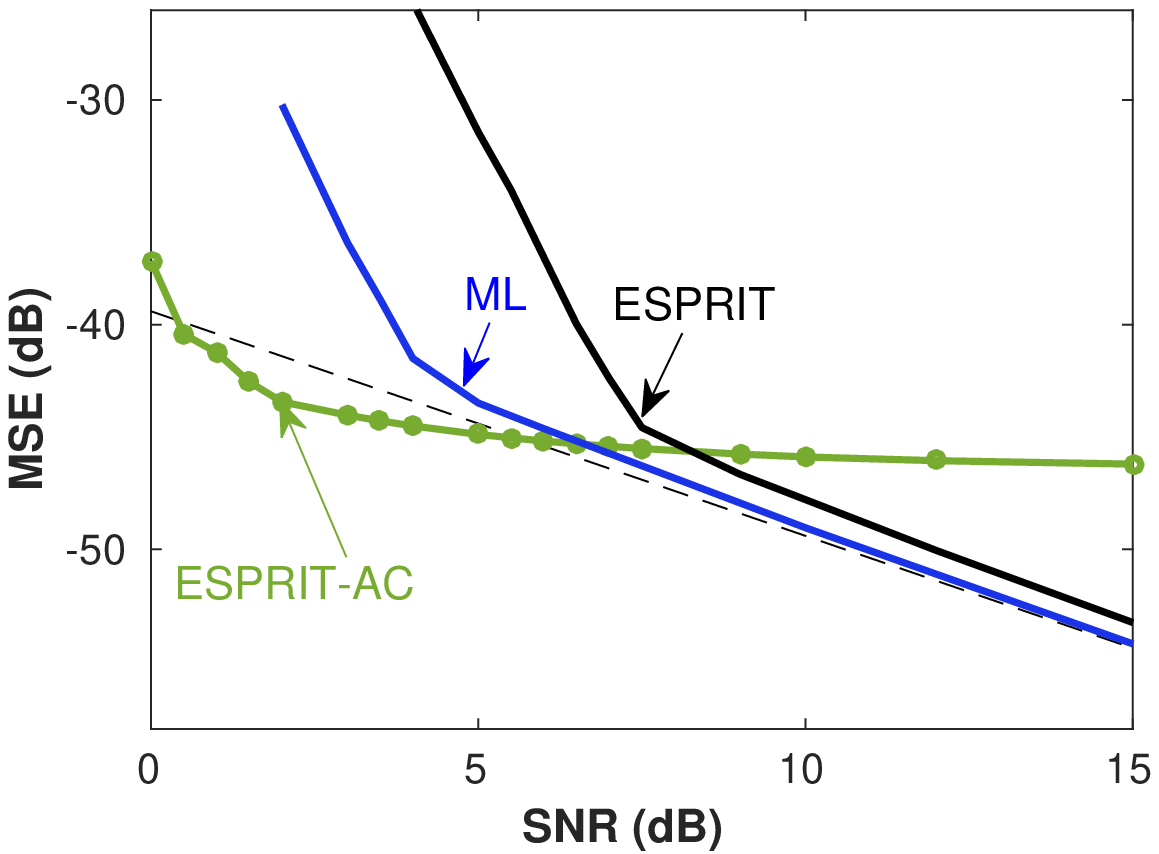}}
\subfloat[Bias]{\includegraphics[width = 1.4in,trim={0 0 0 
0.01cm},clip]{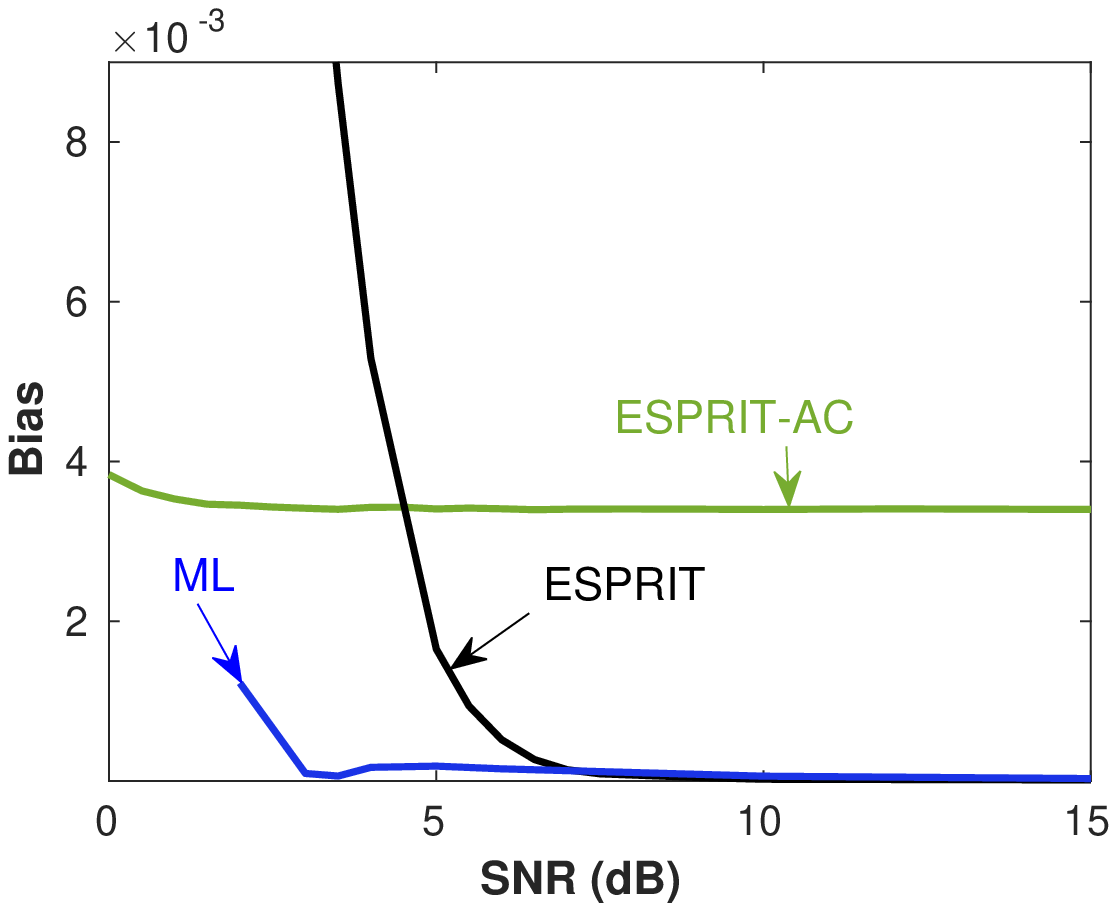}}
\caption{MSE and bias of ML, ESPRIT, and ESPRIT-AC for the well-known two-sinusoid example ($f_1 = 0.52$, $f_2 = 0.5$, $\phi_1 - \phi_2 = 0$).  ESPRIT-AC has a lower threshold than ML but the bias is high.}
\label{fig:zero-pad-compare}
\end{figure}

 
Fig.~\ref{fig:hist-esprit}(a) shows the histogram of $\hat{f}_2$ (based on $50$k trials) obtained for an SNR=$5\,$dB (well below ESPRIT's threshold).  As expected, there are many outliers.  In sharp contrast, the histogram of $\hat{f}_2$ obtained using ESPRIT-AC shows a much smaller variance, but is clearly biased.  For this SNR, the bias was found to be $0.0034$, which may be deemed as acceptable.  However, this bias remains more or less constant even for high SNRs and dominates the MSE, as can be seen from Fig.~\ref{fig:zero-pad-compare}.
Fig.~\ref{fig:zero-pad-compare}(a) shows the overall MSE (sum of MSE for $f_1,\, \ldots, \, f_p$) vs.\ SNR, whereas Fig.~\ref{fig:zero-pad-compare}(b) shows the average bias of $f_1$ and $f_2$ vs.\ SNR.  It is clear from Fig.~\ref{fig:zero-pad-compare}(a) that ESPRIT-AC has a threshold that is lower than that of MLE.  The lower threshold of ESPRIT-AC is directly related to the significantly smaller spread, which is a consequence of the autocorrelation assumption.  However, the unacceptable bias, especially above threshold (Fig.~\ref{fig:zero-pad-compare}(b)), makes this method unusable in its current form.  We propose to carry out additional processing to reduce this bias.  As described in the next section, we use conventional ESPRIT in the high SNR region, and switch to ESPRIT-AC and incorporate additional processing for lower SNRs.  Our previously proposed $\Gamma_\beta$-based method \cite{vishnu-csr-2020} is quite effective for making this switch.

\section{Proposed Method}
The first step in the proposed method is obtaining the initial estimate $\bhf_{\textrm{\tiny init}}$ using conventional ESPRIT.  Since $\bhf_{\textrm{\tiny init}}$ is nearly optimal when the SNR is above ESPRIT's threshold, it is important that it be retained and further refined.  But neither the SNR nor the threshold is known in practice.  In \cite{vishnu-csr-2020} we proposed a method for estimating both the SNR and threshold, and introduced a parameter $\Gamma_{\beta} = \mbox{\textit{Estimated SNR}} - \mbox{\textit{Estimated Threshold}}$.  The results presented there indicate that the parameter $\Gamma_{\beta}$ can be used reliably for deciding whether or not to retain $\bhf_{\textrm{\tiny init}}$.  This parameter is defined as,
\begin{equation}
\Gamma_{\beta} = 10\log_{10}\left[\frac{\hat{\lambda}_p -
	\hat{\sigma}^2}
{M\beta \hat{\sigma}^2}\right]
\end{equation}
where $\hat{\sigma}^2 = \frac{1}{M-p} \sum_{k=p+1}^M \hat{\lambda}_k$ is the estimated noise variance and $\hat{\lambda}_k$, $k=0,1,\ldots, M$ are the eigenvalues of $\bhR_{xx}$ arranged in descending order.

The parameter $\beta$ depends on $M$ and $N$.  Since the total signal power from the eigenvalues of $\bR_{xx}$ equals $P_s = \sum_{i=1}^p \left(\lambda_i - \sigma^2\right)/M$, we can consider $(\hat{\lambda}_p - \hat{\sigma}^2)/M$ as the power contribution of weakest signal component.  That is, we can view $\beta$ as the \textit{minimum ratio between signal power associated with the weakest signal component and noise variance} that guarantees no outlier frequency estimates.  If $M$ and $N$ are known \textit{a priori}, one can fix the value of $\beta$ using any simulation example.  In the case of DoA estimation, the value of $\beta$ depends on $M$ and $K$, where $M$ is the number of antenna elements and $K$ is the number of snapshots.  An approximate expression for $\beta$ has been derived in \cite{vishnu-phd} using a combination of classical asymptotic theory (where $M$ is finite and $K \rightarrow \infty$) and ideas from Random Matrix Theory (where both $M$ and $K$ approach infinity at the same rate).  For $M = 18$, $N = 25$, the required value is $\beta = 0.72$.

If $\Gamma_\beta \leq 0$, we need to first employ ESPRIT-AC, i.e., estimate the frequency $\bhf_{\mbox{\tiny zp}}$ from the zero-padded data. 
This leads us to the question, ``Is $\bhf_{\mbox{\tiny zp}}$ a better estimate?''  To answer it we again use the $\Gamma_\beta$ checking, but this time on the zero-padded data.  That is, now we check whether  $\Gamma_{\beta, \,\mbox{\tiny zp}} > 0$ or not.  If $\Gamma_{\beta, \,\mbox{\tiny zp}} > 0$, i.e., the estimated SNR is above the estimated threshold for the zero-padded data, then we decide that $\bhf_{\mbox{\tiny zp}}$ is a non-outlier estimate; a subsequent gradient descent step is enough to refine it.  Gradient descent leads us to the closest local minimum of cost function about the initial estimate.  The likelihood cost function $L(\beff)$ is given by \cite{kaySpectralBook} 
\begin{equation}
L(\beff) = \bx^H (\bI-\bS\left(\bS^H\bS\right)^{-1}\bS^H)\bx
\label{eq:ML}
\end{equation}
where $\textbf{S}=[\be_1\,\be_2\,\ldots\,\be_p]$, $\be_k=[1\,e^{j2\pi f_k}\,\ldots\, e^{j2\pi(N-1) f_k}]^T$.

\begin{figure}[t!]
\begin{center}
\begin{footnotesize}
  \begin{tikzpicture}[baseline={([yshift=-0.5cm] current bounding box.north)}]
  
  \node (node1) [block,text width= 0.14\linewidth]  {{\color{blue} ESPRIT $\hat{\beff}_{\textrm{\tiny init}}$}};
  
  \node (node2) [decision, text width= 0.15\linewidth, aspect=1.6, below=2mm of node1]  {{\color{blue}$\Gamma_\beta > 0$ ?}};
  
  \draw[->,thick] (node1) edge  (node2);
  \node [draw,circle,minimum size=0.7cm, fill=red!10!white][below=5mm of node2](nodeOut) {$\hat{\beff}_{\textrm{\tiny init}}$};
  \draw[->,thick,black] (node2) -- node[left,near start] {\scriptsize Yes}(nodeOut);

  \node (node5) [block,text width= 0.2\linewidth][right=12mm of node1]{{\color{blue}ESPRIT-AC\\ $\hat{\beff}_{\textrm{\tiny zp}}$}};
  
  \draw[->,thick,black] (node2.east) |- node [near start, right] {\scriptsize No} (node5);
  
  \node (node8) [decision, text width= 0.17\linewidth] [below left=5mm and -5mm of node5.south] {{\color{blue} $\Gamma_{\beta,\mbox{\tiny zp}} > 0$ ?}};
  
  \draw[->,thick,black] (node5.south) -- (node8);
  
  \node (node13) [block,text width= 0.12\linewidth][right=3mm of nodeOut]{{\color{blue}Gradient \\ descent}};
  
  \draw[->, thick, black] (nodeOut) -- (node13);
  
  \node[draw, circle, minimum size=0.5cm,fill=red!10!white, above=1.8mm of node13] (node12) {$\hat{\beff}_{\mbox{\tiny zp}}$};
	
	\draw[->,thick,black] (node8.west) -| node [near start, above] 
	{\scriptsize Yes} (node12);
	
	\node (final1) [below=3mm of node13]{$\hat{\beff}$};
	\draw[->,thick,black] (node13) -- (final1);
	
	\draw[->,thick,black] (node12) -- (node13);
	
	\node (node10) [block,text width= 0.2\linewidth, right=15mm of node13]{{{\color{blue}Remove and\\ re-estimate\\ $\check{\beff}_{\mbox{\tiny zp}}$}}};
	    
	\draw[->,thick,black] (node8.east) -| node [near start, above] {\scriptsize No} (node10);
	\draw[->,thick,black] (node10) -- (node13);

  \end{tikzpicture}\\
\end{footnotesize}
\caption{The proposed algorithm.}
\label{fig:algorithm}
\end{center}
\end{figure}
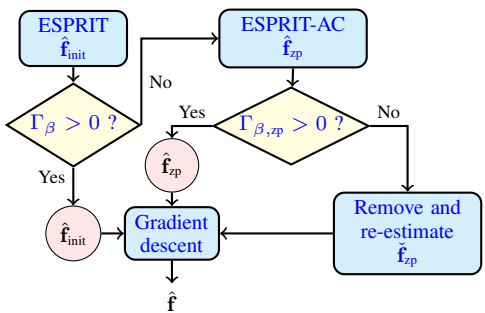

If $\Gamma_{\beta,\,\mbox{\tiny zp}} \leq 0$, we conclude that $\hat{\beff}_{\mbox{\tiny zp}}$ is a poor estimate needing improvement.  This leads to the ``remove and re-estimate'' block.  The principle of removing some components to facilitate better estimation of the remainder is a general one; an example of its use in frequency estimation is the RELAX algorithm of Li and Stoica \cite{li-stoica-1996}.  This method, while still helpful in reducing bias, can fail to resolve closely spaced components; moreover, the bias can be significant if the modes are not well-separated \cite{ying-sabharwal-moses-2000}.  In our implementation, we take advantage of our observation that the maximum improvement is seen for $p=2$ when using ESPRIT-AC followed by gradient descent; for higher values of $p$ these improvements start to diminish.  Hence, for $p>2$, we only consider all possible subsets of $p-2$ sinusoids for removal.  Thus the filtered data contains only two sinusoids at this stage, whose frequencies are re-estimated using ESPRIT-AC.  These are then combined with the other $p-2$ estimates and further refined.  These steps are captured in Fig.~\ref{fig:algorithm}.  The overall algorithm has a much lower computational burden than the MLE (which can be intractable even for $p=4$).  Moreover, for all the illustrative examples considered in this paper (up to $p=5$), the resulting thresholds were found to be lower than MLE's.


The ``remove and re-estimate'' part of the algorithm is initialized with $\check{\beff}_{\mbox{\tiny zp}}^0$.  This is obtained by using a gradient-descent procedure on $L(\beff)$ with $\hat{\beff}_{\mbox{\tiny zp}}$ as the initial estimate.
In the $(i+1)^{\mbox{\tiny th}}$ iteration, the method tries to improve upon $\check{\beff}_{\mbox{\tiny zp}}^i$ as follows.  We partition $\check{\beff}_{\mbox{\tiny zp}}^i$ into two sets: one containing $2$ frequencies and the other the remaining $p-2$ ($Q = \phantom{}^p C_2$ possibilities).  An updated estimate is obtained as follows:
\begin{enumerate}
\item \textbf{Remove}: Let $\bS_l$ be of size $N \times (p-2)$, representing one of $Q$ possibilities.  Filter out these $p-2$ sinusoids from the data $\bx$.  The filtered signal $\tilde{\bx}$ is given by,
\begin{equation}
    \tilde{\bx} = (\bI-\bS_l\left(\bS_l^H\bS_l\right)^{-1}\bS_l^H)\bx
    \label{eq:ML-filter}
\end{equation}
This can be interpreted as matrix notch filtering.  The likelihood cost corresponding the filtered data is $\btx^H\btx$.
\item \textbf{Re-estimate}: Since $p-2$ sinusoids have been removed from $\bx$, we are left with only $2$ sinusoids in $\btx$.  Apply ESPRIT-AC on $\btx$ to re-estimate these $2$ frequencies.  These re-estimates are combined with the $p-2$ frequencies (that were removed in the previous step) and further improved using a gradient descent step (to reduce the likelihood cost further).  The resultant estimate is denoted by $\tilde{\beff}_{\mbox{\tiny zp}}^l$.
\end{enumerate}
We now have a set of frequency estimates $\Omega = \left\lbrace \tilde{\beff}_{\mbox{\tiny zp}}^1, \ldots, \tilde{\beff}_{\mbox{\tiny zp}}^Q \right\rbrace$.  The one with the minimum cost is taken as $\check{\beff}_{\mbox{\tiny zp}}^{i+1}$, i.e.,
\begin{equation}
\check{\beff}_{\mbox{\tiny zp}}^{i+1} = \argmin_{\beff \in \Omega} L(\beff)
\end{equation}
If $L(\check{\beff}_{\mbox{\tiny zp}}^{i+1}) \geq L(\check{\beff}_{\mbox{\tiny zp}}^i)$, then the iteration is stopped.  Next, $\check{\beff}_{\mbox{\tiny zp}}^i$ is used as the initial guess to a gradient-descent routine, whose output yields the final estimate $\bhf$.

We now demonstrate the effectiveness of the removal and re-estimation part of the proposed method with the help of an illustrative example.  Consider the following three sinusoid example: $N=25$, $f_1=0.35$, $f_2=0.5$, $f_3=0.52$, $\phi_1=0$, $\phi_2=\pi/4$, $\phi_3=0$, $|v_1|=1$, $|v_2|=0.5$, $|v_3|=0.53$.  Even for the noiseless case, ESPRIT-AC gives an estimate that is an outlier, i.e., $\hat{\beff}^0_{\textrm{zp}} = [0.3354,\, 0.3594,\, 0.5136]$.  Using this as an initial guess to a gradient descent routine, we get $\check{\beff}^0_{\textrm{zp}}=[0.3177,\, 0.351,\, 0.5105]$, i.e., it is still an outlier with a likelihood cost of $0.7313$.  On the other hand, filtering out $0.351$ and applying ESPRIT-AC results in $[0.4982,\, 0.5225]$.   Appending these values and using $[0.351,\,0.4982,\, 0.5225]$ as the initial estimate to the gradient descent step results in the true frequencies.  This shows that blind zero-padding can yield estimates with large error for $p > 2$.  However, the removal and re-estimation block compensates for the error and gives better estimates.  For $p=2$ gradient descent is enough to bring down the bias.  As a general remark, the proposed algorithm gave the true frequencies as estimates in the noiseless case for all the simulation examples that we have tried.

In all the simulations done so far, we observed that no more than one iteration was required.  The remove/re-estimate block is the most computationally intensive part of the algorithm.  Fortunately, it is used only in a small fraction of the cases (see Table~\ref{rel_freq}).  The simulation results presented in the next section further showcase its effectiveness.  It is important to note that the proposed removal and re-estimation idea is general and can be used by methods other than just ESPRIT-AC.  That is, in the ``remove'' stage, we can partition the frequencies into two sets containing $k$ and $p-k$ components and proceed to the ``re-estimate'' stage.

\section{Simulation Results}
\label{sec:simulation_results}
We present results of the proposed algorithm when applied to the two-sinusoid example of Sec.~\ref{sec:motivation}, and also for sinusoids with \textit{random amplitudes, frequencies, and phases} ($p=3,4,5$; for $p=2$, only $[\phi_1,\;\phi_2]$ was made random).

\begin{figure}[b!]
  \centering \subfloat[MSE]{\includegraphics[width = 
  1.56in]{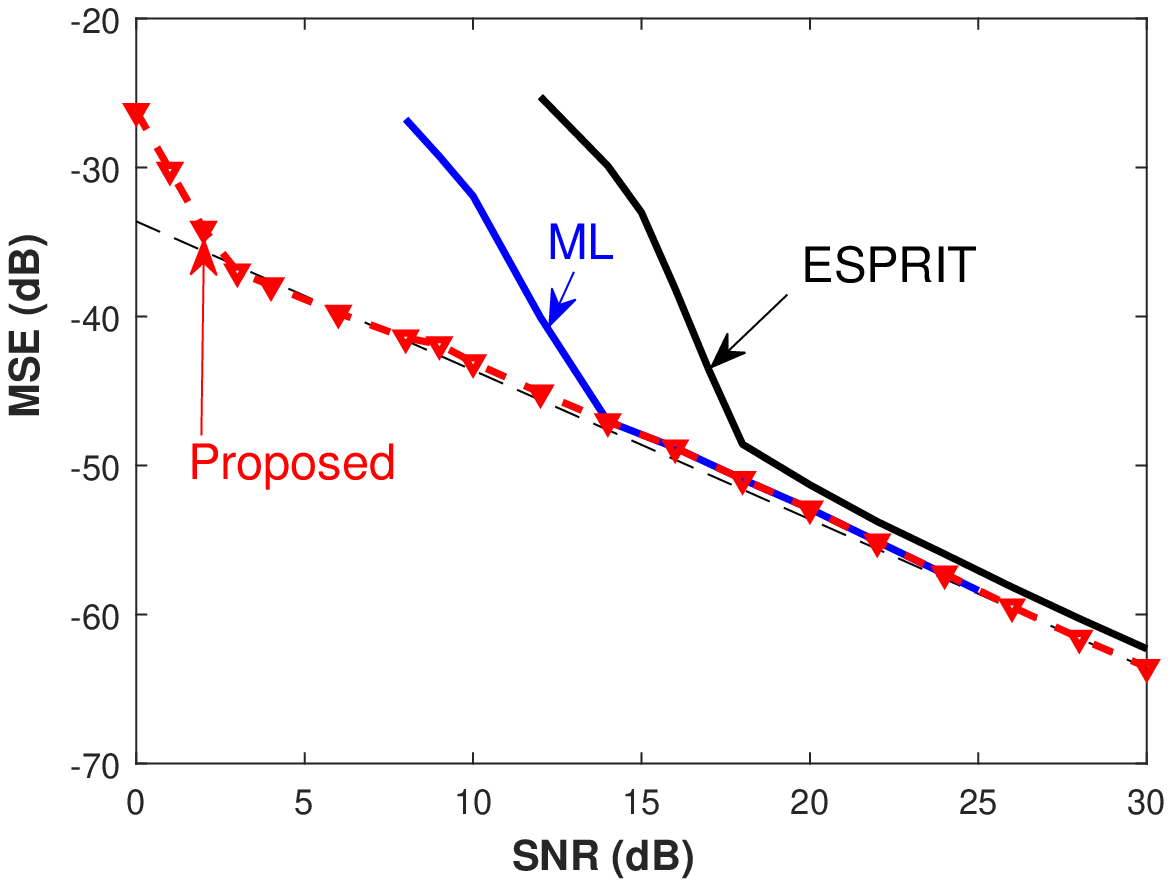}}$\,$
      \subfloat[Bias]{\includegraphics[width = 
      1.5in]{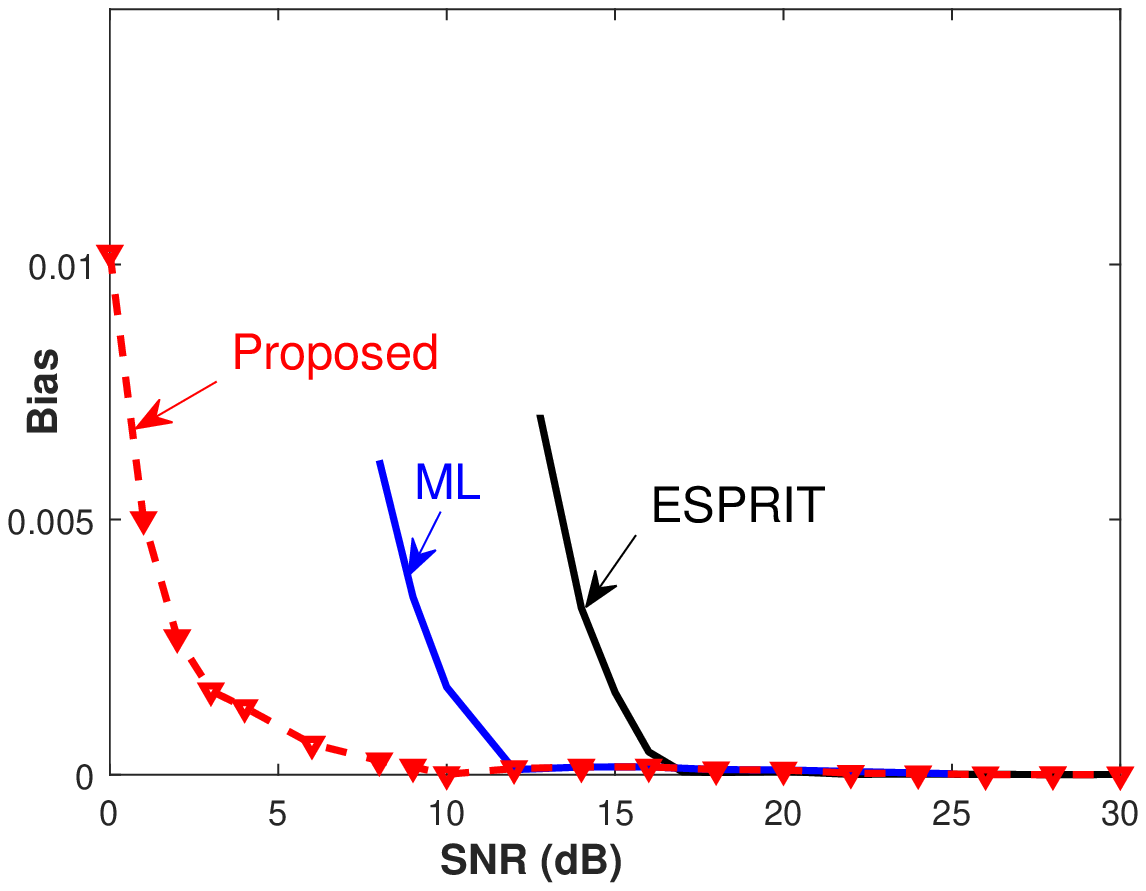}} 
  \caption{Two sinusoids example with random phase, i.e., $\phi_1,\,\phi_2 \sim {\cal U}[0, 2\pi)$ in each trial, with $|v_1| = |v_1| = 1$, $[f_1\;\;f_2]=[0.5\;\;0.52]$.}
\label{fig:random-phase}
\end{figure}
Data containing sinusoids with random parameters were generated as follows.  Both $v_i$ and $f_i$ were chosen randomly ($1000$ realizations): for each $1 \leq i\leq p$, $|v_i| \in {\cal U}[0.5, 1]$, $\phi_i \in {\cal U}[0, 2\pi)$ and $f_i \in {\cal U}[0, 1)$'s, with no two adjacent frequencies getting closer than $\frac{1}{2N}$.  For each random parameter setting we obtained $\bhf$ using $50$ noisy trials, leading to total of $50$k realizations.

\begin{figure}[t!]
  \centering \subfloat[Three 
  sinusoids]{\includegraphics[width=1.9in]{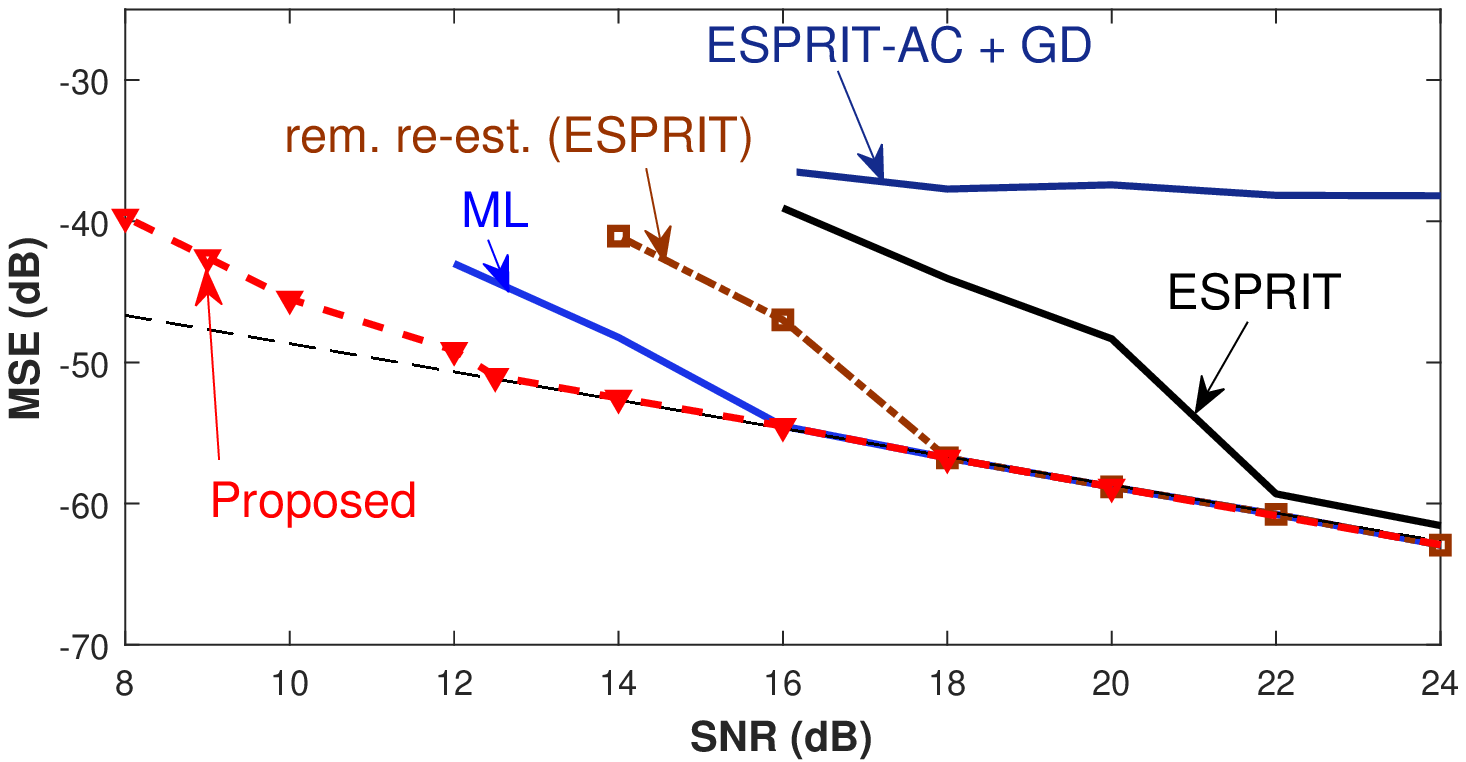}}$\,$
      
    \subfloat[Four 
    sinusoids]{\includegraphics[width=1.53in]{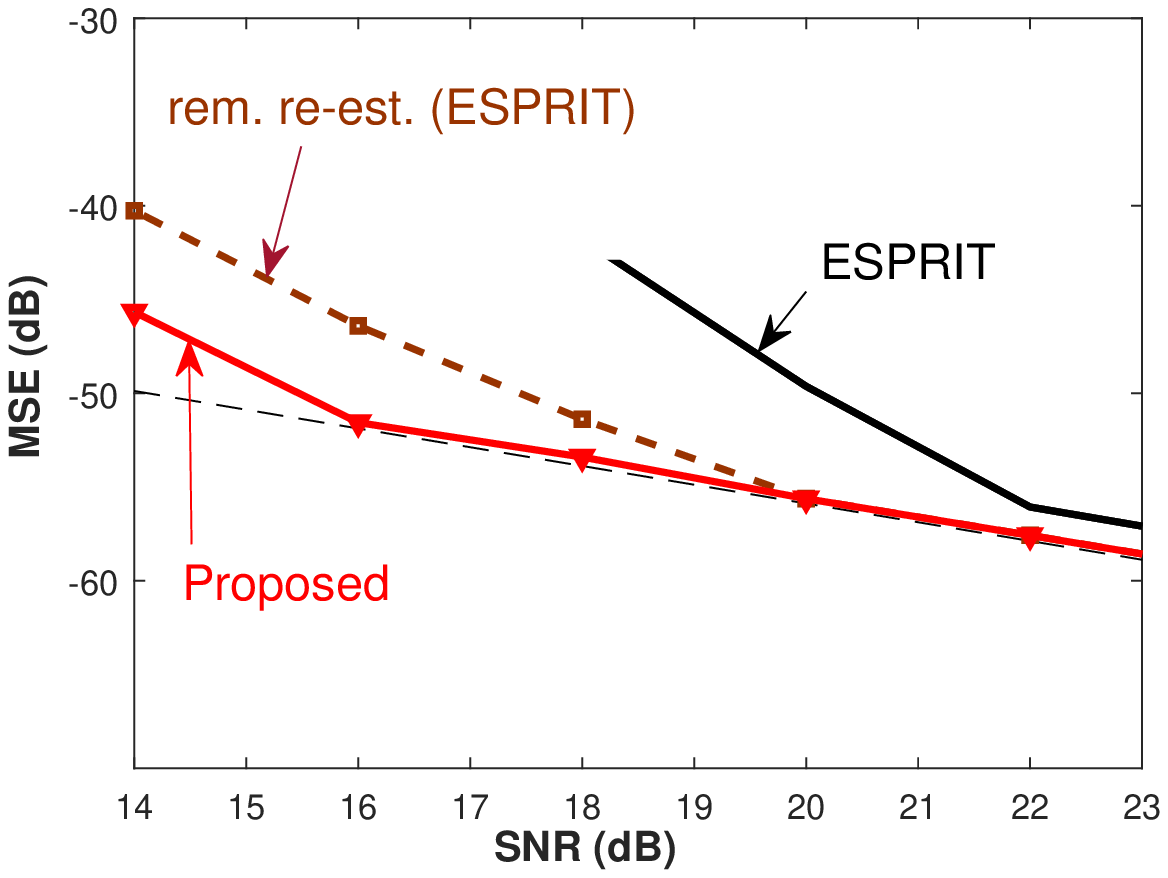}}$\,$
    \subfloat[Five 
    sinusoids]{\includegraphics[width=1.5in]{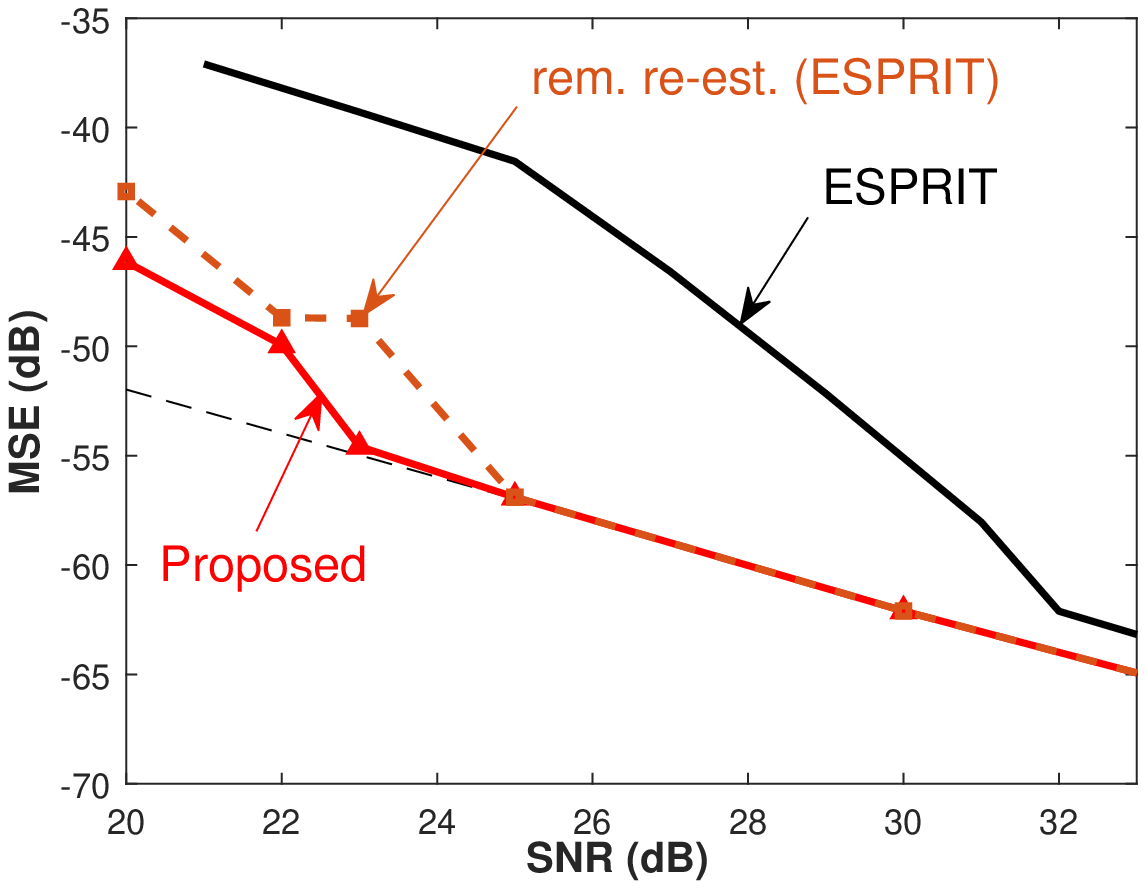}}
  \caption{MSE plots for three, four and five sinusoids, with random amplitudes, frequencies, and phases.}
\label{fig:three-four-sin-mse}
\end{figure}

For $p=2$ with random phase difference, the proposed method's threshold is lower by $10\,$dB compared to ML (Fig.~\ref{fig:random-phase}(a)), and by $4\,$dB for $p=3$ (Fig.~\ref{fig:three-four-sin-mse}(a)).  For $p=4$, the four-dimensional coarse search for the initial guess is computationally too burdensome.  For example, if we choose $500$ points per dimension for the coarse grid search, the number of likelihood evaluations needed will be $\phantom{1}^{500}C_4 = 2.573 \times 10^9$ (for $p=4$) and $\phantom{1}^{500}C_5 = 2.55 \times 10^{11}$ (for $p=5$); this is the reason why the ML curves are absent in Fig.~\ref{fig:three-four-sin-mse}(b) and (c).  Nevertheless one can infer that ML has a higher threshold for $p=4$ also via the following example: $\beff = [0.0526,\;0.0749,\;0.1044,\;0.5299]$ and $16\,$dB SNR (which is the threshold value for the proposed method).
The method in \cite{vishnu-csr-2020} yields $\bhf = [0.0593, 
\;0.1112, \;\textcolor{red}{0.4569}, \;0.5303]$, which is an 
\textit{outlier estimate} (Fig.~\ref{fig:three-four-sin-mse}(b)), 
with likelihood cost $L = 0.5981$.  On the other hand, if we 
initialize gradient descent with $\beff$, we get $\bhf = [0.0570, 
\;0.0704, \;0.1089, \;0.5300]$; crucially, \textit{the likelihood 
cost is higher}, i.e., $L = 0.6515$.  Hence one can easily deduce 
that ML will also give an outlier estimate and that its threshold 
must be higher than $16\,$dB.  For $p=5$, the threshold of the 
proposed method is $23\,$dB.  The following example shows that the 
MLE's threshold is greater than $24\,$dB.  For this SNR the other 
parameters were: phases = $[4.4136,\; 2.4121,\; 0.1956,\; 2.8692,\; 
1.7556]$, amplitudes = $[0.6681,\; 0.5261,\; 0.7700,\; 0.6905,\; 
0.9992]$ and $\beff = [0.3305,\; 0.3536,\; 0.3828,\; 0.7868,\; 
0.8239]$.  When gradient descent was initialized using the true 
frequencies, the estimates were $[0.3340,\; 0.3677,\; 0.3859,\; 
0.7869,\; 0.8241]$, with likelihood cost $0.0816$.  The proposed 
method also gives the same estimate (and hence has the same 
likelihood value).  When the ``remove and re-estimate'' part of the 
algorithm is applied to ESPRIT, it yields $[0.3360,\; 0.3788,\; 
\textcolor{red}{0.5864},\; 0.7872,\; 0.8240]$, which has a larger 
peak absolute error.  The corresponding likelihood value is 
\textit{lower}, i.e., $0.0761$.  Hence the ML estimate's likelihood 
value has to be less than or equal to $0.0761$.  Its peak absolute 
error will also have to be larger than the estimate given by the 
proposed method.  From this we conclude that MLE's threshold has to 
be greater than or equal to $24\,$dB.  Thus, for the $p=5$ case 
considered here, the proposed method has a lower threshold and 
smaller peak absolute error.

\begin{figure}[b!]
	\centering
	\subfloat[]
	{\includegraphics[width = 1.8in]{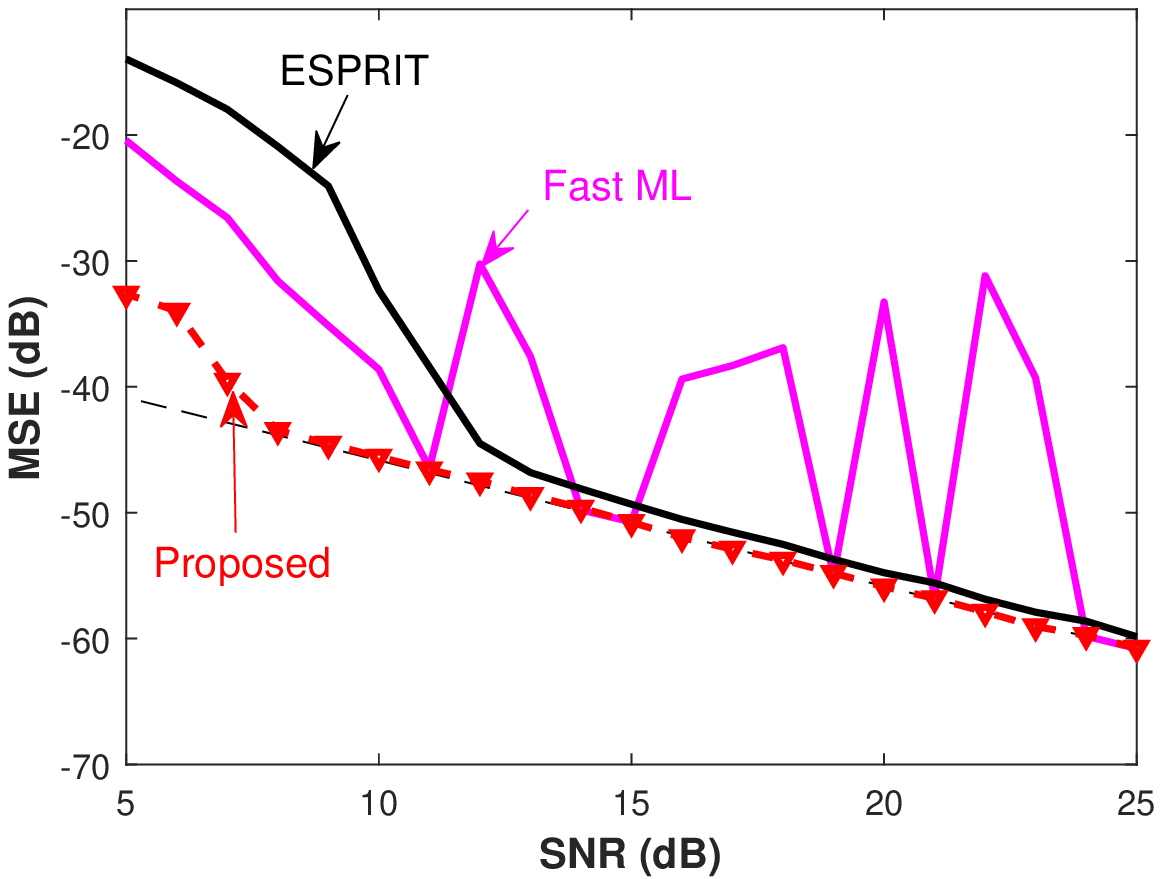}}\\
	\subfloat[]
	{\includegraphics[width = 1.8in]{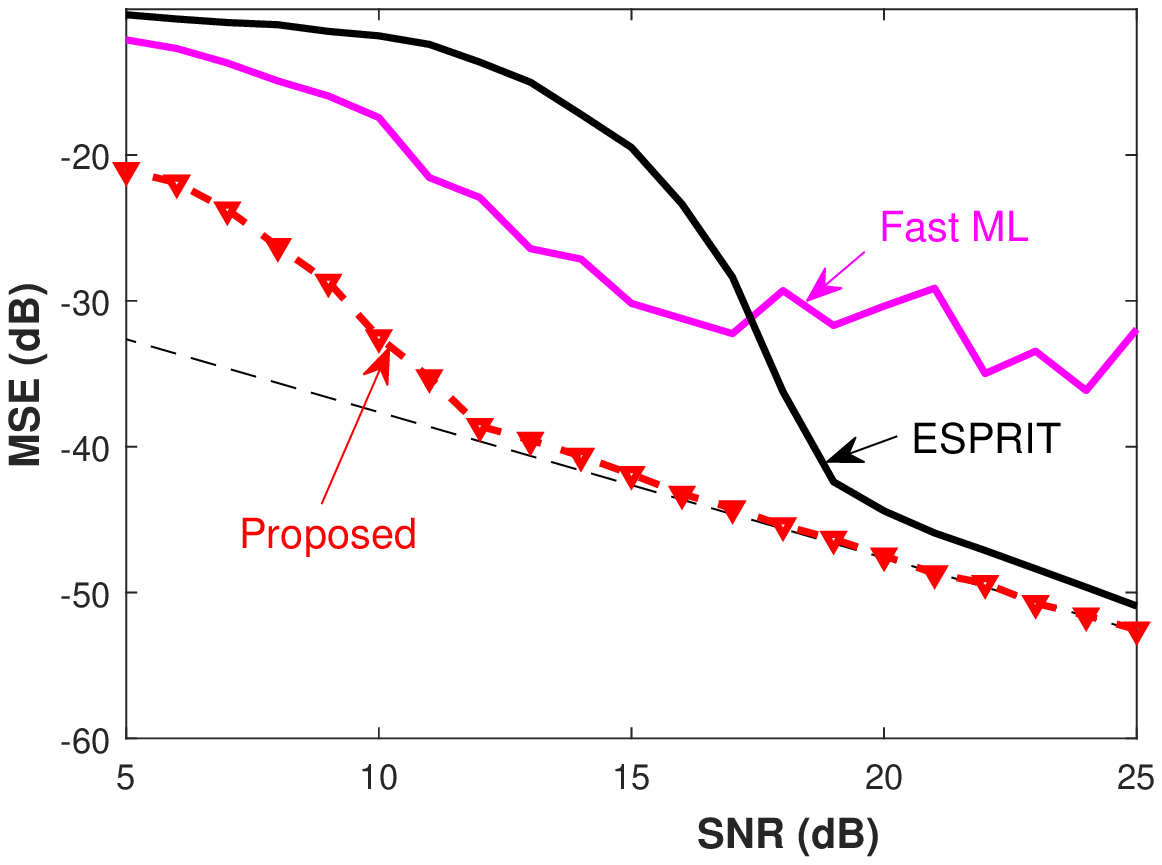}}
	\caption{ Poor performance of FastML method for the three sinusoids examples with $N=25$, $f_1=0.52$, $f_2= 0.5$, $f_3=0.3$, $|v_1| = |v_2| = |v_3| = 1$ for two different starting phases: (a) $\phi_1=0$, $\phi_2=\pi/4$, $\phi_3=0$, and (b) $\phi_1=0$, $\phi_2=\pi/2$, $\phi_3=\pi$.  MSE was obtained from $2k$ trials.
	}
	\label{fig:egFastML}
\end{figure}
In this context, it is worth recalling the number of likelihood computations that were required for $p=4$ and $p=5$.  The Fast ML method \cite{tufts1993fastMlPaper} was proposed mainly to address this computational burden.  It is reduced by replacing the computationally burdensome $p$-dimensional coarse search by $p$ one-dimensional searches.  Unfortunately, this method not only fails to guarantee ML (or even near-ML) performance but also performs unacceptably poorly in some cases, as the following examples show.

Consider the following three-sinusoid case: $N=25, f_1=0.52, f_2=0.5, 
f_3=0.3, |v_1|=|v_2|=|v_3|=1$. Fig.~\ref{fig:egFastML} shows the MSE 
plots for two different phase differences: (a) $\phi_1=0, 
\phi_2=\pi/4, \phi_3=0$, and (b) $\phi_1=0, \phi_2=\pi/2, 
\phi_3=\pi$.  These plots show that Fast ML is even poorer than the 
ESPRIT method for these chosen cases. The jagged nature of the curve 
in Fig.~\ref{fig:egFastML}(a) is due to the method's sensitivity to 
the initial frequency choices: even a small difference in the 
initialization may give rise to large variations in the final 
estimates; this is due to the highly nonlinear nature of the 
likelihood function.  This is best illustrated in the noiseless case: 
if the initial guesses are chosen to be $0.3$, $0.4$, and $0.5$, the 
final estimates turn out to be the true frequencies, i.e., $0.3$, 
$0.5$, and $0.52$.  On the other hand, even if one of the initial 
estimates is slightly perturbed, i.e., if the initial estimates are 
chosen as \textcolor{red}{$0.31$}, $0.4$ and $0.5$, then the 
resulting final estimates are quite poor, viz., $0.3010, 0.3163, 
0.51$.  Clearly, these examples demonstrate that the Fast ML method 
has unacceptably poor performance, despite being computationally 
tractable. Note that our proposed method does not face these 
initialization issues of Fast ML and is consistently better, as 
evident from Fig.~\ref{fig:egFastML}.

The effectiveness of the proposed method for SNRs below ML's threshold in the two-sinusoid case with random $\phi_1-\phi_2$ can be seen when $\beff = [0.5,\;0.52]$ and $\bphi = [0,\;4.3069]$ at $8\,$dB SNR.  Both ESPRIT's estimate ($\bhf = [\textcolor{red}{0.0518},\;0.5109]$, $L = 2.8126$) and that of ML ($\bhf = [0.5110, \;\textcolor{red}{0.9608}]$, $L = 2.6259$) are outliers.  On the other hand, the proposed method gives $\bhf = [0.5034, \;0.5205]$, having smaller $\|\beff - \hat{\beff}\|_\infty$, even though the likelihood cost $L = 2.9661$ is higher.

In the above examples, despite the SNR being well below ML's threshold, the proposed method yields estimates that are closer to the true value albeit with higher likelihood costs.  This brings out the effectiveness of the zero-padding scheme and subsequent processing.

\begin{figure}[h!]
\centering \subfloat[MSE]{\includegraphics[width = 1.55in]
{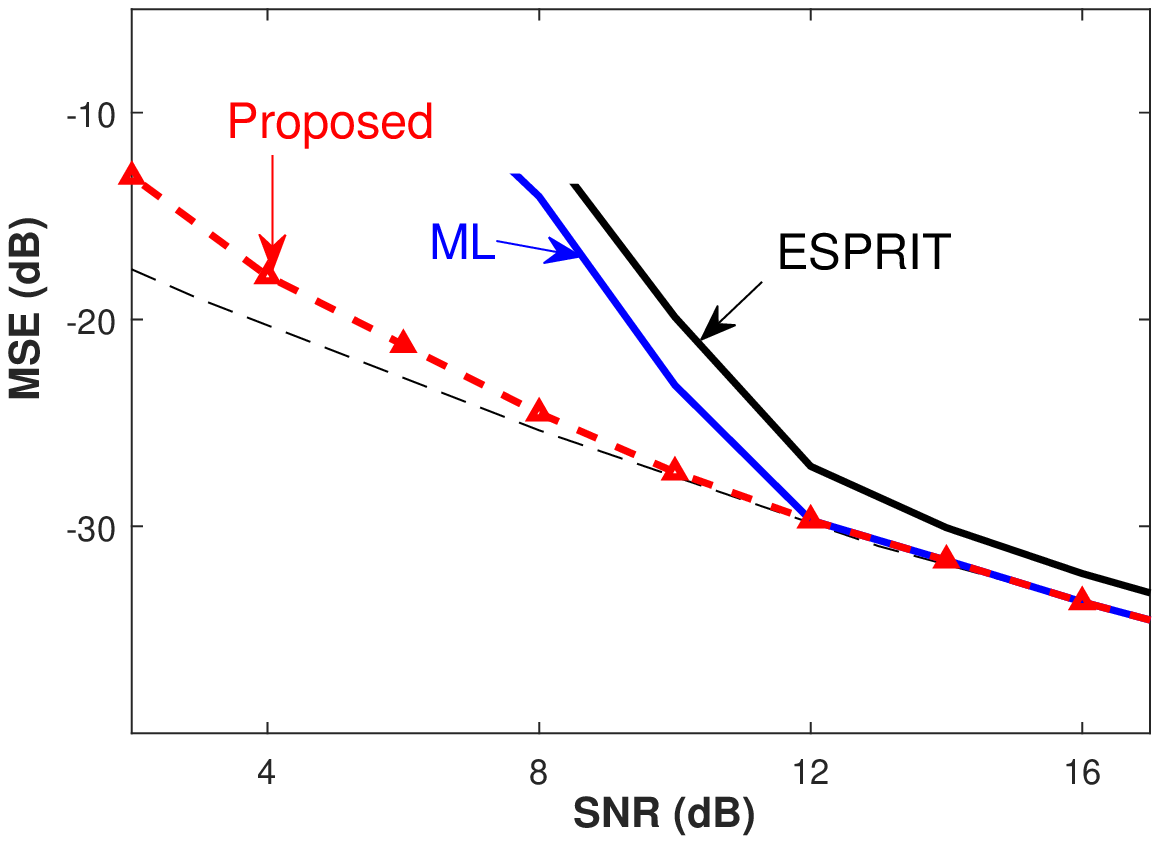}}$\,$
\subfloat[Bias]{\includegraphics[width = 1.64in]
{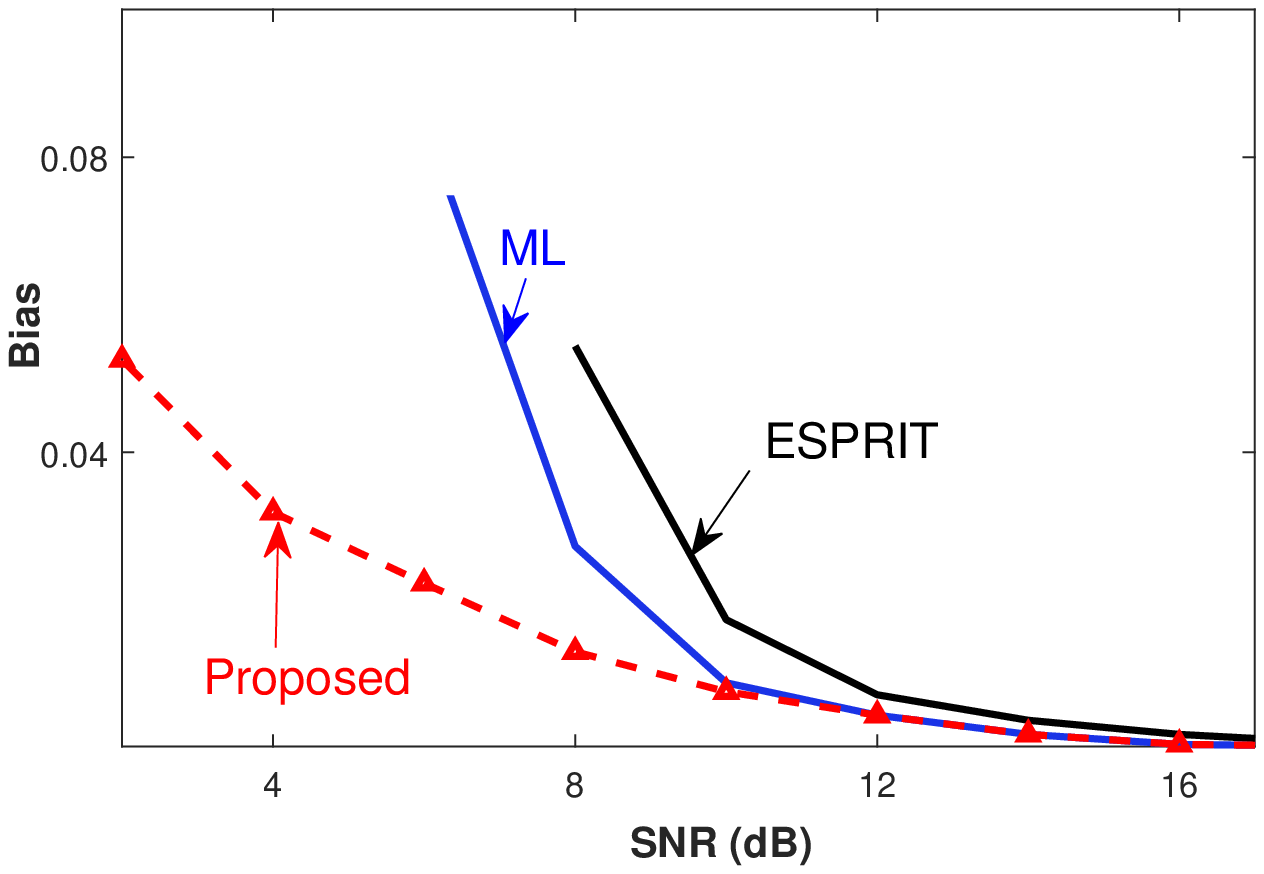}} 
\caption{DoA estimation example: $p=2$, array size = $10$, number of snapshots = $10$, $\phi_1 = 35^{\circ}$, $\phi_2 = 37^{\circ}$.}
\label{fig:doa}
\end{figure}

The proposed method was applied to the DoA estimation example given in \cite{Shaghaghi-Vorobyov-2015}.  The parameters were: $p=2$, array size $M=10$, number of snapshots $L=10$, $\phi_1 = 35^{\circ}$, $\phi_2 = 37^{\circ}$.  The results are given in Fig.~\ref{fig:doa}.  For this example also the proposed method has a lower threshold than ML, doing better by $2\,$dB.  It is noteworthy that below threshold value the increase in variance for this particular example is far more gradual than the usual sudden increase.

Fig.~\ref{fig:three-four-sin-mse}(a) illustrates that the performance of ESPRIT-AC + gradient-descent is far from adequate.  On the other hand, ESPRIT-AC followed by the ``remove and re-estimate'' step contributes significantly to the improvement in performance.  Fig.~\ref{fig:three-four-sin-mse} also shows that removal and re-estimation step applied to ESPRIT-AC is more effective than applying it on ESPRIT.

While carrying out removal and re-estimation, it is quite possible that more than one choice of the partition into $2$ and $p-2$ sinusoids will lead to the same final estimate.  For example, for $p=4$ and SNR=$20\,$dB, the true frequencies were $\beff = [0.4505,\, 0.4755,\, 0.64,\, 0.7735]$.  The initial estimate was $\check{\beff}^0_{\textrm{zp}} = [\textcolor{red}{0.6028}, 0.4786, 0.6358, 0.7739]$.  This is an outlier estimate caused by the presence of $0.6028$.  Removing the frequency components $0.4786$ and $0.7739$ in the filtering step leads to the final estimate $\check{\beff}^1_{\textrm{zp}}=[0.451, 0.4726, 0.6396, 0.7744]$, which is better than $\check{\beff}^0_{\textrm{zp}}$.  Removing the components with frequencies $0.6358$ and $0.4786$ also results in the same $\check{\beff}^1_{\textrm{zp}}$.  On the other hand, removing combinations of the ``unsuccessful frequency" $0.6028$ does not improve the result.  In the current version, all possible subsets having $p-2$ frequencies are considered and the final estimate is chosen.  A computationally more efficient algorithm that is optimal or nearly so is being investigated. 

The relative frequencies of how often the various parts of the algorithm were exercised (based on 50k trials) for the $p=2, 3$ examples is given in Table~\ref{rel_freq}.  It is seen that the computationally most expensive part, i.e., the removal and re-estimation step, is used only in a small fraction of the trials.  For the three-sinusoid example with SNR = $18\,$dB, for those cases for which $\Gamma_{\beta} < 0$, the average ``minimum distance between frequencies'' was $0.0246$; the maximum value of this minimum distance was $0.0372$.  Hence these cases contain closely spaced sinusoids.


\begin{table}[h!]
  \centering
  \begin{tabular}{p{16mm}ccc|ccc}
    \toprule
    &  \multicolumn{3}{c|}{Two sinusoids} & \multicolumn{3}{c}{Three sinusoids}\tabularnewline\cline{2-7}
    SNR (dB) $\rightarrow$ &  $6$ & $10$ & $14$ & $14$ & $16$ & $18$\Tstrut\\\midrule
    ESPRIT & \textsf{0.371} & \textsf{0.700} &
    \textsf{0.858} & \textsf{0.977} & \textsf{0.984} & \textsf{0.990}\\[0.5ex]
    ESPRIT-AC & \textsf{0.629} & \textsf{0.300} &
    \textsf{0.142} & \textsf{0.016} & \textsf{0.013} & \textsf{0.010}\\[0.5ex]
    ESPRIT-AC+ Rem/re-est & - & - &
    - & \textsf{0.007} & \textsf{0.003} & \textsf{0}\\\bottomrule
    \end{tabular}
    \caption{How often the various parts of the algorithm are used, as a function of SNR, averaged over 50k trials for the two- and and three-sinusoid examples.}
    \label{rel_freq}
  \end{table}


\balance
\section{Conclusion}
In this paper we proposed a method for sinusoidal frequency estimation that yields a threshold that is lower than that of the MLE.  For the examples considered herein, the improvements over MLE is up to $10\,$dB, which is very significant.  Moreover, the bias is either equal to or lower than that of the MLE for the SNR range considered.  The key to the improvement lies in the fact that estimates produced by ESPRIT-AC have lower variance, but are biased.  Subsequent processing, such as the $\Gamma_\beta$-based checking and the ``remove and re-estimate'' block, have contributed to threshold SNRs that are lower than MLE's.  The proposed removal and re-estimation, when applied to other methods (e.g, ESPRIT), leads to a lowering of their thresholds as well.


\bibliographystyle{IEEEtran}
\bibliography{IEEEabrv,ref,references}

\end{document}